\documentclass[conference, 10pt]{IEEEtran}
\IEEEoverridecommandlockouts
\usepackage{cite}
\usepackage{amsmath,amssymb,amsfonts}
\usepackage{algorithmic}
\usepackage{graphicx}
\usepackage{textcomp}
\usepackage{xcolor}
\def\BibTeX{{\rm B\kern-.05em{\sc i\kern-.025em b}\kern-.08em
    T\kern-.1667em\lower.7ex\hbox{E}\kern-.125emX}}
\usepackage[letterpaper, left=1.05in, right=1.05in, bottom=1in, top=0.75in]{geometry}

\usepackage{booktabs}

\usepackage[english]{babel}

\usepackage{multirow}

\usepackage{url}

\usepackage{numprint}
\npdecimalsign{.}

\usepackage{acronym}
\newacro{cots}[COTS]{Commercial Off The Shelf}   
\newacro{iiot}[IIoT]{Industrial Internet of Things}   
\newacro{iot}[IoT]{Internet of Things}     
\newacro{it}[IT]{Information Technology}
\newacro{ot}[OT]{Operation Technology}
\newacro{ip}[IP]{Internet Procotol}  
\newacro{hmi}[HMI]{Human Machine Interface}
\newacro{plc}[PLC]{Programmable Logic Controller}  
\newacro{cnc}[C\&C]{Command \& Control}  
\newacro{ics}[ICS]{Industrial Control System} 
\newacro{ids}[IDS]{Intrusion Detection System}
\newacro{scada}[SCADA]{Supervisory Control And Data Acquisition}  
\newacro{swat}[\textit{SWaT}]{\textit{Secure Water Treatment}}  
\newacro{lstm}[LSTM]{Long Short-Term Memory}
\newacro{cps}[CPS]{Cyber-Physical System}
\newacro{wsn}[WSN]{Wireless Sensor Networks}
\newacro{arima}[ARIMA]{Auto-Regressive Integrated Moving Average}
\newacro{dos}[DoS]{Denial of Service}    
\newacro{uv}[UV]{Ultraviolet}    
\newacro{ro}[RO]{Reverse Osmosis}    
\newacro{svm}[SVM]{Support Vector Machine}    
\newacro{sssp}[SSSP]{Single Stage Single Point}
\newacro{ssmp}[SSMP]{Single Stage Multi Point}
\newacro{mssp}[MSSP]{Multi Stage Single Point}
\newacro{msmp}[MSMP]{Multi Stage Multi Point}
\newacro{mtu}[MTU]{Master Terminal Unit}
\newacro{uf}[UF]{Ultra Filtration}
\newacro{ocsvm}[\textit{OCSVM}]{\textit{One Class Support Vector Machine}}
\newacro{stamp}[STAMP]{Scalable Time series Anytime Matrix Profile}
\newacro{rnn}[RNN]{Recurrent Neural Network}
\newacro{dnn}[DNN]{Deep Neural Network}
\newacro{gan}[GAN]{Generative Adversarial Network}
\newacro{cnn}[CNN]{Convolutional Neural Network}
\newacro{s317}[\textit{S317}]{\textit{SUTD Security Showdown (S3) 2017}}

\begin{document}

\title{Intrusion Detection in Binary Process Data: \\
Introducing the \textit{Hamming}-distance to \textit{Matrix Profiles}\thanks{This is a preprint of a submission accepted at the 6th  IEEE International workshop on
Communication, Computing, and Networking in Cyber Physical Systems (CCNCPS 2020). Please cite as:
\textit{SD Duque Anton, HD Schotten: Intrusion Detection in Binary Process Data: Introducing the Hamming-distance to Matrix Profiles,
in Proceedings of the 20th IEEE International Symposium on a World of Wireless, Mobile and Multimedia Networks (IEEE WoWMoM 2020), 2020.}}
}

\author{\IEEEauthorblockN{1\textsuperscript{st} Simon D Duque Anton}
\IEEEauthorblockA{\textit{Intelligent Networks Research Group} \\
\textit{German Research Center for AI}\\
Kaiserslautern, Germany\\
\textit{simon.duque\_anton@dfki.de}}
\and
\IEEEauthorblockN{2\textsuperscript{nd} Hans D Schotten}	
\IEEEauthorblockA{\textit{Intelligent Networks Research Group} \\
\textit{German Research Center for AI}\\
Kaiserslautern, Germany\\
\textit{hans\_dieter.schotten@dfki.de}}
}

\maketitle

\begin{abstract}
The digitisation of industry provides a plethora of novel applications that increase flexibility and reduce setup and maintenance time as well as cost.
Furthermore,
novel use cases are created by the digitisation of industry,
commonly known as \textit{Industry 4.0} or the \textit{Industrial Internet of Things},
applications make use of communication and computation technology that is becoming available.
This enables novel business use cases,
such as the digital twin,
customer individual production,
and data market places.
However,
the inter-connectivity such use cases rely on also significantly increases the attack surface of industrial enterprises.
Sabotage and espionage are aimed at data,
which is becoming the most crucial asset of an enterprise.
Since the requirements on security solutions in industrial networks are inherently different from office networks,
novel approaches for intrusion detection need to be developed.
In this work,
process data of a real water treatment process that contains attacks is analysed.
Analysis is performed by an extension of \textit{Matrix Profiles},
a motif discovery algorithm for time series.
By extending \textit{Matrix Profiles} with a \textit{Hamming}-distance metric,
binary and tertiary actuators can be integrated into the analysis in a meaningful fashion.
This algorithm requires low training effort while providing accurate results.
Furthermore,
it can be employed in a real-time fashion.
Selected actuators in the data set are analysed to highlight the applicability of the extended \textit{Matrix Profiles}.
\end{abstract}	


\section{Introduction}
\label{sec:intro}
The introduction of \textit{Industry 4.0},
also referred to as the \ac{iiot},
enables an abundance of novel use cases~\cite{3gpp2017,Haller.2008}.
These use cases in turn introduce new business cases,
meaning industrial value generation is changed.
Customer-individual processing with minimal delay and digital twins are examples of the scenarios that can be implemented because of digitisation in industry.
However,
since these novel use cases rely heavily on intercommunication and computation,
the network structures of industrial enterprises,
the \ac{ot} networks,
are changing.
When \ac{scada} systems were first introduced in the 1970's,
they were meant to control industrial devices in a pre-defined,
non-flexible fashion.
Furthermore,
the networks were physically separated from public networks and highly application specific~\cite{Igure.2006}.
These features limited the surface for an attacker.
\ac{cots} products for industrial application and a focus on interconnectivity drastically increase the attack surface,
making a focus on intrusion detection necessary~\cite{Duque_Anton.2017a}.
Since the requirements for industrial \acp{ids} are different than requirements for office \ac{it} environments,
novel approaches have to be developed.
\textit{Powers et al.}~\cite{Powers.2015} as well as \textit{Iturbe et al.}~\cite{Iturbe.2017} discuss the different operational conditions of \ac{it} and \ac{ot} environments.
Generally,
\ac{ot} networks are operated for longer periods,
i.e. decades,
constantly with little possibility to update or change systems.
\acp{ids} must not affect the process,
since availability is the highest rated requirement. \par
The contribution of this work consists of:
\begin{itemize}
\item Integration of the \textit{Hamming}-distance~\cite{Hamming.1950} into the \textit{Matrix Profile}-algorithm~\cite{Yeh.2016a}, and
\item detection of attacks in a real process environment by analysing actuator information.
\end{itemize}
The remainder of this work is structured as follows.
Section~\ref{sec:related_work} presents related work to industrial intrusion detection.
The data set used to evaluate the usefulness of the extended \textit{Matrix Profiles} is discussed in Section~\ref{sec:presenting_the_data_set},
the standard and extended \textit{Matrix Profiles} are introduced in Section~\ref{sec:introduction}.
Section~\ref{sec:evaluation} evaluates the performance.
A discussion of the algorithms and results is presented in Section~\ref{sec:discussion}.
This work is concluded in Section~\ref{sec:conclusion}.

\section{Related Work}
\label{sec:related_work}
This section presents scientific works related to intrusion detection in industrial environments,
with a focus on process data analysis.
As discussed in Section~\ref{sec:intro},
the relevance of this topic increases.
Hence,
it is widely addressed in research.
\textit{Schneider and B\"{o}ttinger} use \textit{autoencoders} for detecting human-based attacks on an industrial environment~\cite{Schneider.2018},
provided by the \textit{iTrust,
Centre for Research in Cyber Security,
Singapore University of Technology and Design}~\cite{iTrust.2017}.
The \textit{autoencoders} are capable of detecting the attacks in the data set called \ac{s317} in an unsupervised fashion.
Another data set provided by this institute is analysed by \textit{Goh et al.}~\cite{Goh.2017}.
\acp{rnn} are employed to detect the attacks introduced into the industrial environment.
The same data set is evaluated in this work.
Furthermore,
this data set has been analysed by means of \textit{Matrix Profiles} already,
with a focus on the sensor data~\cite{Duque_Anton.2019e, Duque_Anton.2018c}.
A counter has been introduced that was capable of not only detecting outliers,
but also detecting similar attacks that occur rarely in comparison to motifs occurring often~\cite{Duque_Anton.2019g}.
Generally,
the \ac{swat} data set has been widely regarded in research.
\textit{Inoue et al.} analyse it with \acp{dnn} as well as \textit{ocsvm}~\cite{Inoue.2017}.
Similarly,
\textit{Kravchik and Shabtai} employ \acp{cnn}~\cite{Kravchik.2018}.
\textit{Li et al.} analyse the data set with \acp{gan}~\cite{Li.2019, Li.2018}.
A method with code mutation is presented by \textit{Chen et al.}~\cite{Chen.2018}.
\textit{Lin et al.} develop a graphical model to detect the attacks~\cite{Lin.2018}.
\ac{lstm},
a type of neural network,
is applied in a multi-level approach to detect attacks in a gas pipeline data set by \textit{Feng et al.}~\cite{Feng.2017}.
Means to increase the security of industrial networks are discussed by \textit{Knapp and Langill}~\cite{Knapp.2014}.
However,
a significant disadvantage of neural networks is the immense training data and effort required to build expressive models.
\textit{Yang et al.} introduce detection methods for attacks in power system networks~\cite{Yang.2014}.
A publication summarising the stages and development of \ac{scada} security systems is presented by \textit{Larkin et al.}~\cite{Larkin.2014}.
One task difficult to achieve is detecting attacks that are formerly unknown to the operators.
Learning a model of the normal system state and detecting deviations is a strength of machine learning algorithms,
such as \acp{ocsvm}.
\textit{Maglaras and Jiang} present their application to detecting novel attacks~\cite{Maglaras.2014}.
Industrial communication protocols often do not provide means for authentication and encryption~\cite{Herrero_Collantes.2015}.
Due to their long operation times,
they are still in use,
resulting in the need for additional security measures.
\textit{Gao and Morris}present an approach for signature-based intrusion detection in \textit{Modbus}-networks~\cite{Gao.2014}.

\section{Presenting the Data Set}
\label{sec:presenting_the_data_set}
\begin{figure}
  \includegraphics[width=\linewidth]{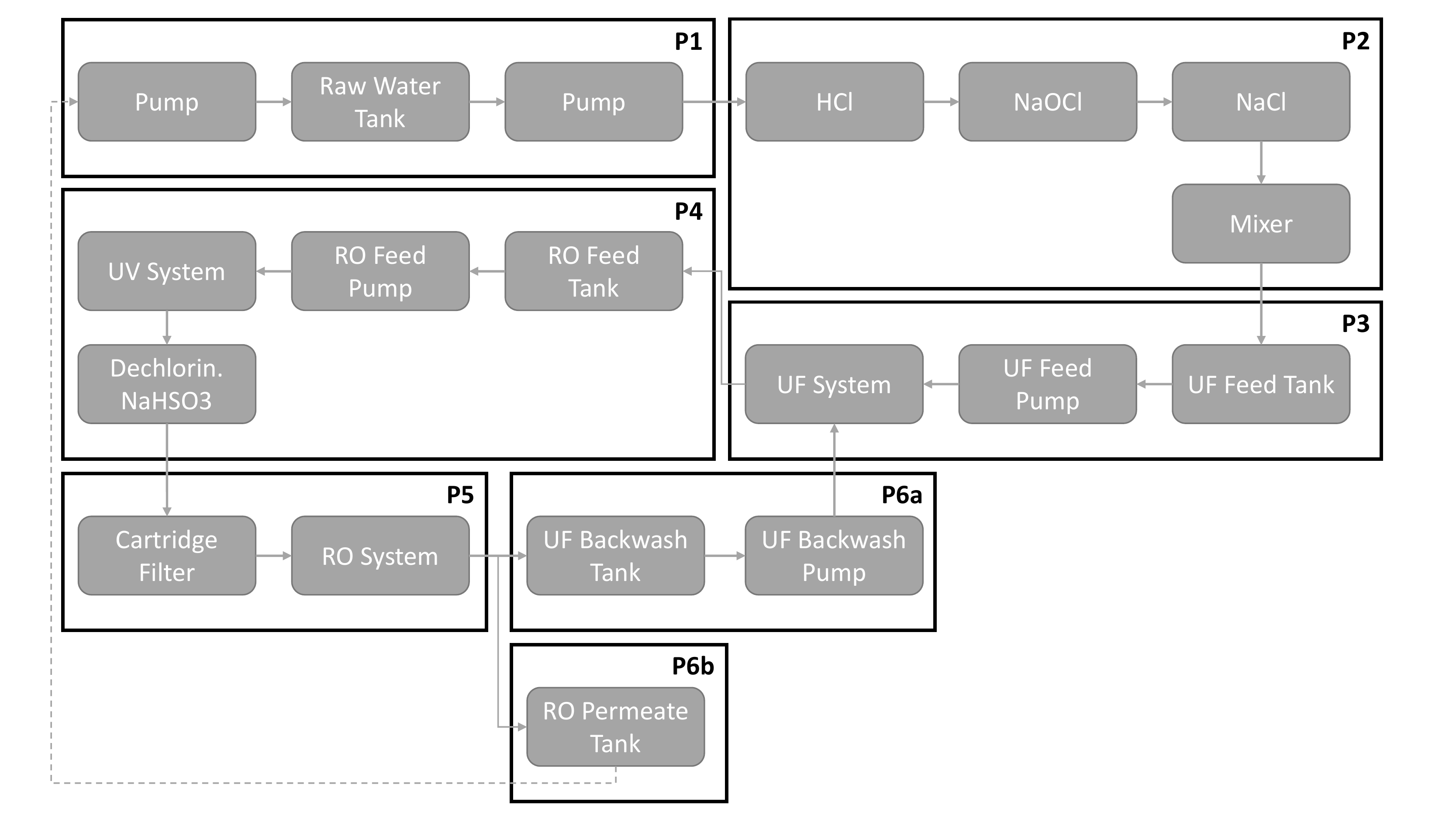}
  \caption{Relation of Sub-Processes}
  \label{fig:process_order}
\end{figure}

\begin{figure}
  \includegraphics[width=\linewidth]{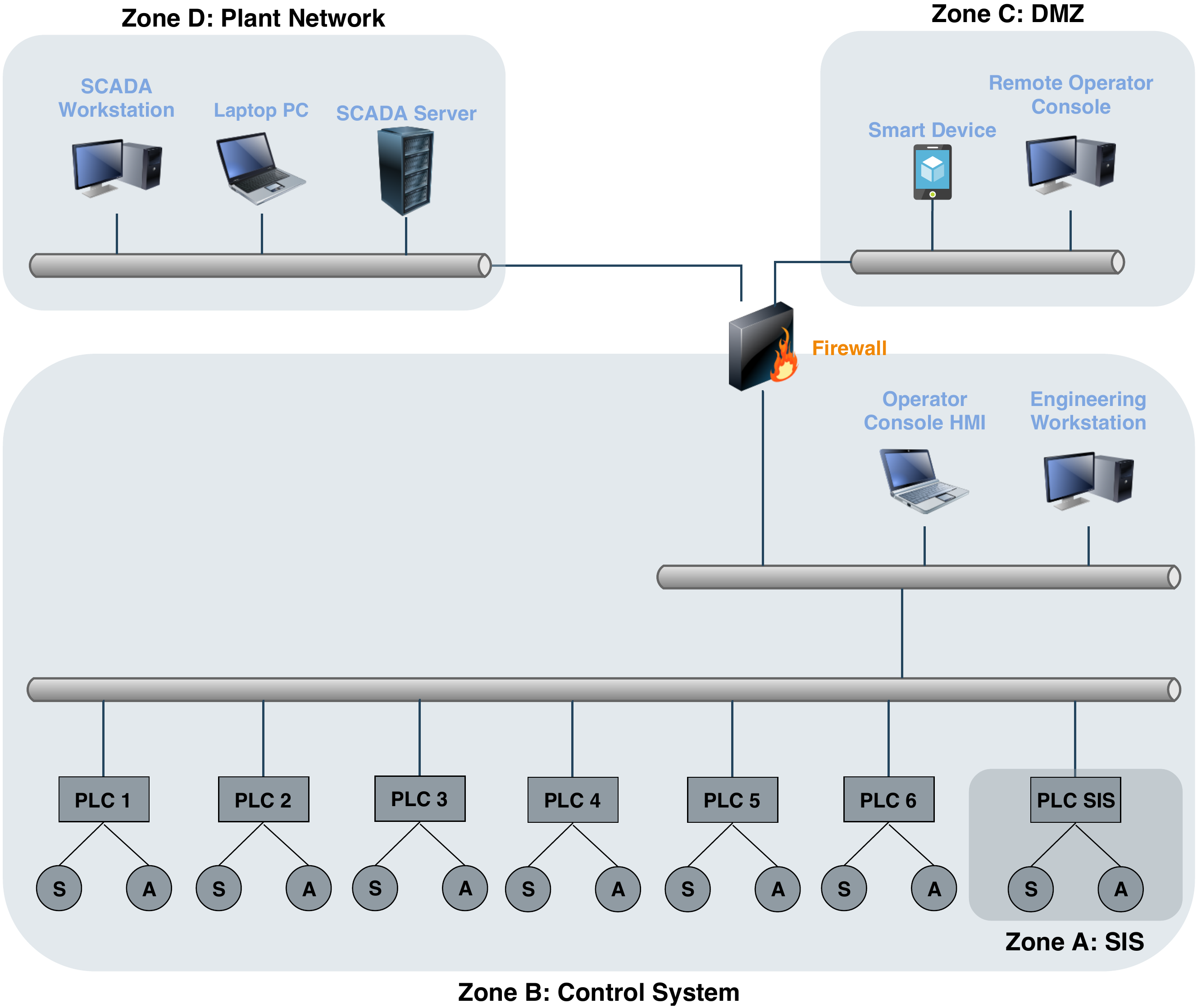}
  \caption{Schematic Overview of the Process Environment}
  \label{fig:nw_structure}
\end{figure}

The data set analysed in this work is created by the \textit{iTrust,
Centre for Research in Cyber Security,
Singapore University of Technology and Design}.
It is named \ac{swat}~\cite{Goh.2016, iTrust.2018}.
Creating data sets for intrusion detection is a crucial,
yet non-trivial task~\cite{Duque_Anton.2019a}.
It has been previously addressed by research,
e.g. by \textit{Schneider and B\"ottinger}~\cite{Schneider.2018}.
The \ac{swat} data set was captured in a water processing facility consisting of real equipment.
A process of water treatment was run for eleven days.
During the first seven days,
only normal operation occured. 
During the last four days,
attacks were introduced.
The attacker was assumed to already have broken the perimeter,
thus directly impacting the \acp{plc}.
Overall,
one of six \acp{plc} was used to control a respective sub-process.
The sub-processes are as follows:
\begin{itemize}
\item \textit{P1}: Raw water storage
\item \textit{P2}: Pre-treatment
\item \textit{P3}: Membrane \ac{uf}
\item \textit{P4}: Dechlorination by \ac{uv} lamps
\item \textit{P5}: \ac{ro}
\item \textit{P6}: Disposal
\end{itemize}
The relation of the sub-processes is shown in Figure~\ref{fig:process_order}.
Each \ac{plc} controls a ring network,
while the \acp{plc} are controlled by a \ac{scada} workstation.
Data is collected by a data historian.
The network structure is depicted in Figure~\ref{fig:nw_structure}.
The raw water is contained in an initial tank and pre-processed.
In a second step,
filtration as well as \ac{uv} light and \ac{ro} treatment are applied.
Then the water is stored in another tank,
given the treatment has resulted in sufficiently clean water.
Otherwise,
the \ac{uv} light and \ac{ro} treatment are repeated.\par
During the four last days of operation,
41 different instances of attacks are introduced into the process.
The attacks are changes in process variables with the malicious intent to disrupt the operation.
Ground truth,
i.e. a labelling known to be correct,
is provided.
Each of the 41 attacks falls into one of four categories,
as introduced by \textit{Goh et al.}~\cite{Goh.2016}:
\begin{itemize}
\item \textit{\ac{sssp}}: Single stage attack on one point in the process, 26 instances in the data set
\item \textit{\ac{ssmp}}: Single stage attack on multiple points in the process, 4  instances in the data set
\item \textit{\ac{mssp}}: Multi stage attack on one point in the process, 2 instances in the data set
\item \textit{\ac{msmp}}: Multi stage attack on multiple points in the process, 4 instances in the data set
\end{itemize}
Some of the attacks,
18 in total,
could not be observed to have an influence on the process.
51 sensors and actuators provided information about the process that could be used to detect the attacks,
a complete list is provided by \textit{Goh et al.}~\cite{Goh.2016}.
The top five sensors and actuators with respect to most attacks aimed at them are listed in Table~\ref{tab:source}.
This table includes the number of attacks and the number of attacks that did not affect the process.
\begin{table}[h!]
\renewcommand{\arraystretch}{1.3}
\caption{Sources of the Attacks}
\label{tab:source}
\centering
\scriptsize
\begin{tabular}{l l l r r}
\toprule
\textbf{Elem} & \textbf{Sub-P} & \textbf{Description} & \textbf{Total} & \textbf{No Ch} \\
P-102 & P1 & Pump (backup) & 3 & 0 \\
P-101 & P1 & Pump & 2 & 0 \\
MV-101 & P1 & Motor valve  & 2 & 0 \\
P-302 & P3 & \ac{uf} feed pump & 2 & 0 \\
P-203 & P2 & Dosing pump & 2 & 0 \\
\bottomrule
\end{tabular}
\end{table}
Each attack had a point in the process on which it was started and a target.
Similar to the starting points,
the top five sensors and actuators that were targets of an attack are listed in Table~\ref{tab:detectable},
together with the number of attacks and the number of attacks that did not influence them.
\begin{table}[h!]
\renewcommand{\arraystretch}{1.3}
\caption{Detectable Points of Attacks}
\label{tab:detectable}
\centering
\scriptsize
\begin{tabular}{l  l  l r r}
\toprule
\textbf{Elem} & \textbf{Sub-P} & \textbf{Description} & \textbf{Total} & \textbf{No Ch} \\
LIT-101 & P1 & Raw water tank level & 7 & \phantom{m} 3 \\
P-101 & P1 & Pump & 2 & 0 \\
LIT-301 & P3 & \ac{uf} feed tank level & 5 & 3 \\
MV-303 & P3 & Motorised valve & 2 & 0 \\
LIT-401 & P4 & \ac{ro} feed tank level & 3 & 1 \\
\bottomrule
\end{tabular}
\end{table}

\section{Intrusion Detection in Industrial Data}
\label{sec:introduction}
This section presents the algorithm used to detect anomalies in industrial process data.
A general assumption is that in an industrial environment,
deviation from expected behaviour is either an attack,
a user error or malfunction and thus worth noting and inspecting.
The employed algorithm is not capable of distinguishing different kinds of anomalies.
In the first subsection,
the general algorithm as well as the application on continuous data are explained.
An extension to employ the algorithm in a meaningful fashion on binary data as well is introduced in the second subsection.

\subsection{Continuous Data}
The \textit{Matrix Profile} algorithm has previously been employed successfully to detect attacks in the data set presented in Section~\ref{sec:presenting_the_data_set}~\cite{Duque_Anton.2019g, Duque_Anton.2019e, Duque_Anton.2018c}.
However,
the algorithm as presented by \textit{Yeh et al.} relies on continuous data.
\textit{Matrix Profiles} were presented by \textit{Yeh et al.} in 2016 as an algorithm for motif discovery~\cite{Yeh.2016a}.
It worked by analysing the sequences of length $m$ starting at each point $t_n$ of the time series and comparing it to each other sequence of length $m$.
The sequences are analysed in a sliding window-fashion.
A distance of the sequences is calculated,
for example the z-normalised distance as described in (\ref{eq:z_norm_dist}).
\begin{equation}\label{eq:z_norm_dist}
\begin{split}
d(x,y) = \sqrt{\sum_{i=1}^{m}{(\hat{x}_{i} - \hat{y}_{i})}^2} \\
\hat{x}_{i} = \frac{x_{i} - \mu_{x}}{\sigma_{x}},\quad \hat{y}_{i} = \frac{y_{i} - \mu_{y}}{\sigma_{y}}
\end{split}
\end{equation}
After applying \textit{Pearson's Correlation Coefficient}~\cite{Benesty.2009} (\ref{eq:pearson})
\begin{equation}\label{eq:pearson}
\begin{split}
corr(x,y) &= \frac{E((x - \mu_x)(y-\mu_y))}{\sigma_x \sigma_y} \\
& = \frac{\sum^{m}_{i=1}x_i y_i - m \mu_x \mu_y}{m \sigma_x \sigma_y},
\end{split}
\end{equation}
where
\begin{equation}\label{eq:mu}
\begin{split}
\mu_x = \frac{\sum_{i=1}^{m} x_i}{m}, \quad \mu_y = \frac{\sum_{i=1}^{m} y_i}{m}
\end{split}
\end{equation}
and
\begin{equation}\label{eq:sigma}
\begin{split}
\sigma_{x}^{2} = \frac{\sum_{i=1}^{m} x_{i}^{2}}{m} - \mu_{x}^{2}, \quad \sigma_{y}^{2} = \frac{\sum_{i=1}^{m} y_{i}^{2}}{m} - \mu_{y}^{2}.
\end{split}
\end{equation}
The Euclidean distance is derived as indicated in (\ref{eq:relation})~\cite{Mueen.2010},
\begin{equation}\label{eq:relation}
\begin{split}
d(x,y) = \sqrt{2m(1-corr(x,y))}
\end{split}
\end{equation}
the resulting distance metric is shown in (\ref{eq:working_formular_dist}).
\begin{equation}\label{eq:working_formular_dist}
\begin{split}
d(x, y) = \sqrt{2m\bigg(1-\frac{\sum_{i=1}^{m} x_{i} y_{i} - m \mu_{x} \mu_{y}}{m \sigma_{x} \sigma_{y}}\bigg)}
\end{split}
\end{equation}
$x$ and $y$  are two distinct time series,
$\mu$ is the respective mean and $\sigma$ the respective standard deviation.
The minimal distances are calculated and stored in a matrix fashion,
which leads to the name \textit{Matrix Profiles}.
If the minimal distance is high in comparison to the minimal distances of other sequences,
the corresponding sequence is an outlier or anomaly in the time series,
since there is no similar sequence contained.
On the other hand,
small minimal distances indicate the presence of at least one similar sequence.

\subsection{Binary Data}
In contrast to sensors in industrial environments that typically produce continuous values,
e.g. the temperature,
pressure or flow volume,
actuators often return a binary value.
This value indicates the operation state of the actuators,
commonly either active or inactive.
The event space is binary,
i.e. either on or off,
which can be represented by boolean values of 0 and 1.
If these values are considered as a time series and the \textit{Matrix Profiles} are computed with the distance metrics provided by the authors,
the calculation often breaks in praxis.
If there are constant values,
or means of 0,
divisions by zero occur.
This characteristic makes the current distance metrics unsuited for binary data.
To bridge this gap,
the \textit{Hamming} distance~\cite{Hamming.1950} is proposed as an additional distance metric for calculation of the \textit{Matrix Profiles}.
The \textit{Hamming} distance $D(x, y)$ is defined ``as the number of coordinates for which x and y are different''~\cite{Hamming.1950}.
Applied to the task of evaluating the distance of binary sensors,
that means a sequence of length $m$ is compared to any other sequence in the time series of the same length $m$ in a sliding window fashion.
The sequences are compared,
the distance is derived by calculating the number of bits that are different between the two sequences for a given position. 
This approach results in a distance which can then be used to compute the \textit{Matrix Profiles} as introduced in the previous subsection.

\section{Evaluation}
\label{sec:evaluation}
This section presents the application of the extended \textit{Matrix Profiles} as presented in Section~\ref{sec:introduction} to the data set introduced in Section~\ref{sec:presenting_the_data_set}.
The application of \textit{Matrix Profiles} to continuous sensor data obtained from the same data set has been successfully evaluated in related works~\cite{Duque_Anton.2019g, Duque_Anton.2019e, Duque_Anton.2018c}. \par
In this work,
a total of three time intervals was selected from the data set and evaluated for attacks.
These intervals contain selected attacks on the top four sources for attacks,
listed in Table~\ref{tab:source}.
The sequence length $m$ was set to \numprint{2000} or \numprint{500}.
Those values were used as preliminary evaluations provided promising results for these values.
In contrast to machine learning-based intrusion detection,
the \textit{Matrix Profiles} algorithm does not require a training phase and training data as such.
Instead,
each sequence is compared to every other sequence.
A general assumption about batch processing is the periodicity of behaviour,
i.e. the process variables repeat themselves over and over again.
In case of water treatment,
a batch of water is introduced to the treatment process,
treated in each stage,
and removed from the environment.
This process is then repeated with the next batch of water,
which creates highly similar values in terms of process control.
This leads to the assumption that events occurring repeatedly are intended,
while events only happening once are malicious or non-intentional.
Furthermore,
one period of normal operation is sufficient to detect normal and anomalous behaviour.
In the course of this work,
several periods of normal operation were employed,
since there might be small deviations. \par
\begin{figure}[!ht]
\centering
\includegraphics[width=\linewidth]{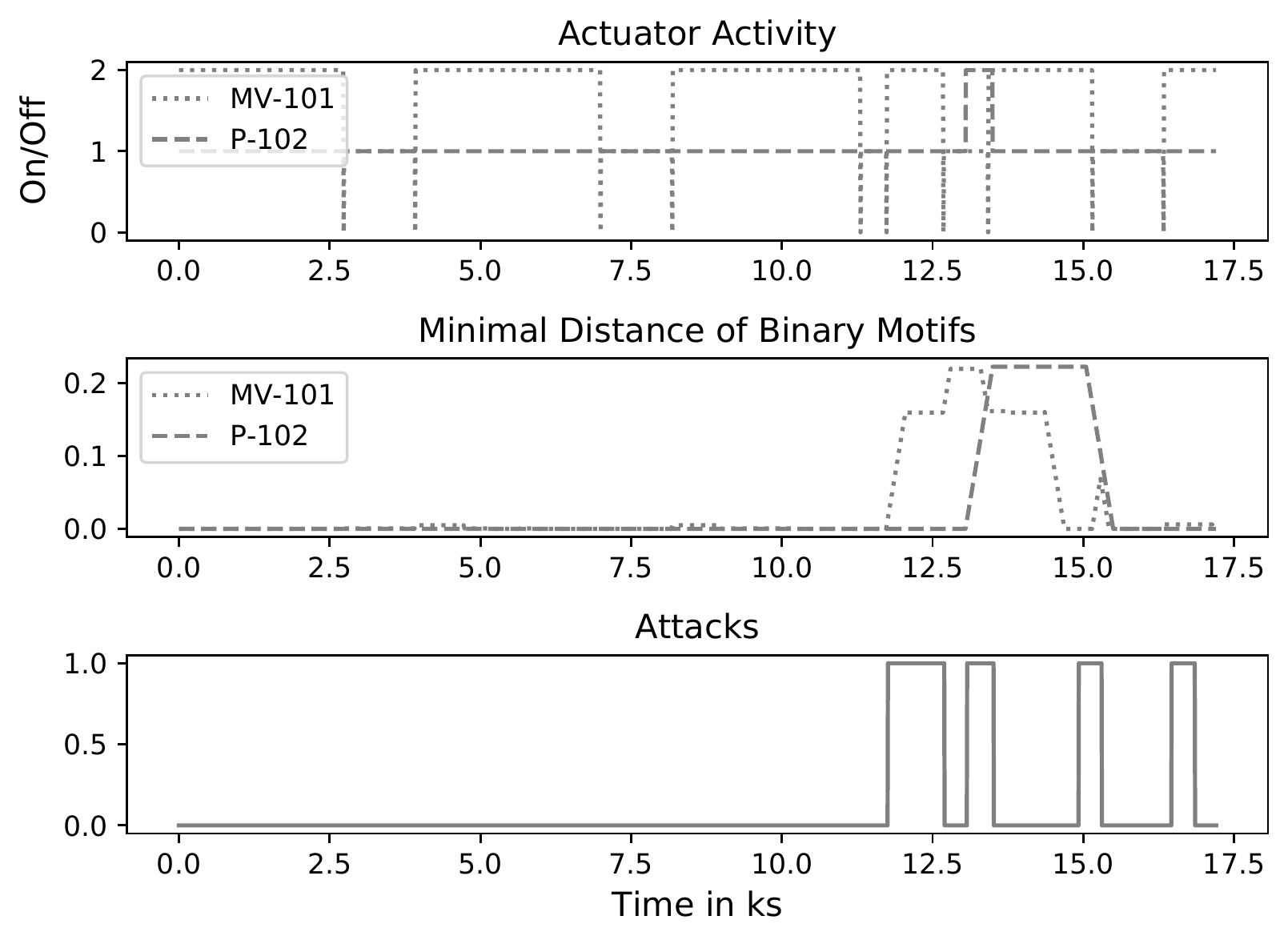}
\caption{\textit{Matrix Profile} of Interval 1 for \textit{MV-101} and \textit{P-102}}
\label{fig:interval1}
\end{figure}

With this in mind,
each interval analysed in this work consists of a period without attacks,
followed by two instances of attacks.
For each interval,
one or two sensors are analysed.
In the first and third interval,
two actuators are evaluated.
In the second interval,
one actuator was analysed.
The first interval starts with the last \numprint{10000} events of the normal data set,
i.e. the seven day period during which no attacks were introduced.
Appended are the first \numprint{7203} events of the malicious data set,
i.e. the four day period during which all attacks occurred.
That means \numprint{10000} events can be considered as training data,
the next \numprint{7203} events contain the attacks,
but they contain normal operation as well.
An overview of the interval is shown in Figure~\ref{fig:interval1}.
The actuator output for the motorised valve \textit{MV-101} controlling water flow to the raw water tank as well as for the backup pump \textit{P-102} pumping water from the raw water tank to the next stage are shown in the top line with the dotted line indicating \textit{MV-101} and the dashed line \textit{P-102}.
For some reason that is not explained by the authors of the data set,
\textit{MV-101} provides tertiary output, 
i.e. zero, 
one,
and two.
In the second line,
the minimal distances are shown,
again with the dotted line indicating \textit{MV-101} and the dashed line \textit{P-102}.
Attacks are indicated in the bottom line.
Overall,
four attacks occur during the interval,
however,
the last two attacks are affecting neither \textit{MV-101} nor \textit{P-102} and are not expected to be discoverable by observing these actuators.
Still,
the third attack aims at the raw water tank level so that it implicitly also affects the valve \textit{MV-101},
leading to a smaller raise in minimal distance around second \numprint{15000}.
Due to the inter-dependability of components,
attacks can be detected by looking at devices that are not target of the given attack.
The first attack unexpectedly opens the motorised valve \textit{MV-101},
leading to an uncontrolled water flow into the raw water tank and potential overflow.
This unexpected behaviour is clearly distinguishable in Figure~\ref{fig:interval1},
meaning this attack can be detected.
The second attack is turning on the backup pump \textit{P-102} increasing the pressure in the pipe from raw water tank to initial stage of treatment.
This could lead to a pipe burst.
Similar to the first attack,
this attack can be detected by the change of minimal \textit{Hamming}-distance of the \textit{Matrix Profile} as the backup pump is expected to remain inactive unless a malfunction requires it to become active.
Figure~\ref{fig:interval1} shows that the minimal distance is continuously increasing with the start of the attack,
as for each sequence after the attack,
more data points are different from known sequences.
Hence,
defining a sensible threshold value allows for early discovery of the attack.
In this case,
a threshold around \numprint{0.1} would easily detect all attacks with no false positives. \par
\begin{figure}[!ht]
\centering
\includegraphics[width=\linewidth]{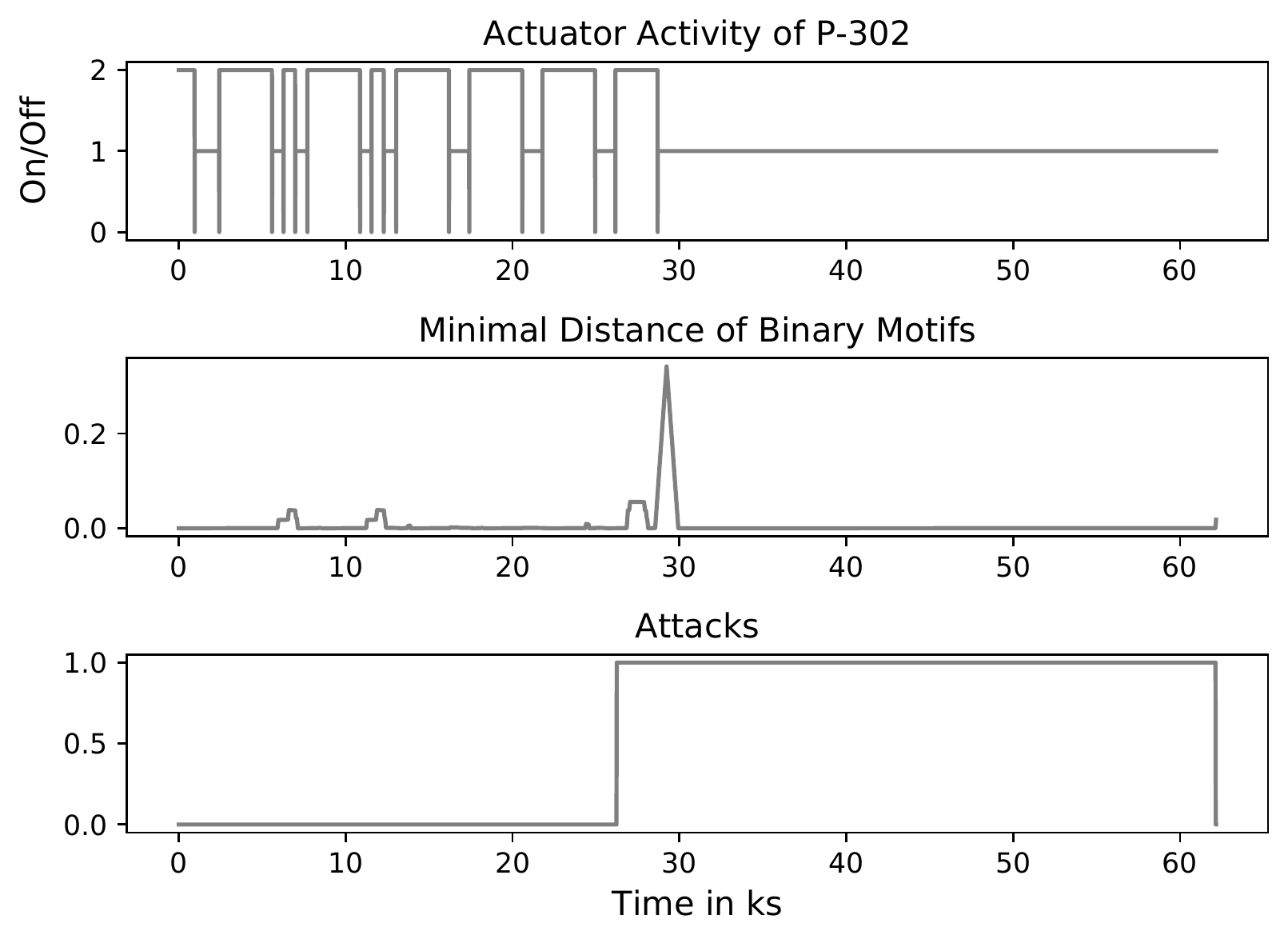}
\caption{\textit{Matrix Profile} of Interval 2 for \textit{P-302}}
\label{fig:interval2}
\end{figure}

Interval 2 is the period on December 31st from 18:00:00 to 23:16:00, 
covering \numprint{62160} events during which two attacks occur.
The actuator under observation is the pump \textit{P-302} that pumps water from the \ac{uf} feed tank into the \ac{ro} feed tank.
As a first attack,
\textit{P-302} is kept on despite of the \ac{ro} feed tank having reached capacity.
As a second attack,
\textit{P-302} is turned off to stop the flow of water.
This behaviour is shown in Figure~\ref{fig:interval2}.
Since the two attacks in this interval are occurring back to back,
there is only one attack-peak visible.
The second attack lasts for several hours,
thus the length.
The first attack leads to a slight peak in the minimal distance,
as it consists of leaving the pump open.
Since the first attack only affects the system briefly,
its characteristic is hardly different than normal operation.
The second attack,
however,
leads to a notable peak in the minimal distance,
clearly marking an anomaly.
Both attacks can be detected automatically,
given an appropriate threshold is selected.
Additionally,
the width of the peak can be used as an indicator of an anomaly. 
The window size $m$ was set to \textit{2000} for analysis of this interval.
A value of \textit{500} led to constant minimal distance of 0,
presumably since a window size smaller than a period of operation does not contain relevant information,
i.e. the length and structure of one period.\par
\begin{figure}[!ht]
\centering
\includegraphics[width=\linewidth]{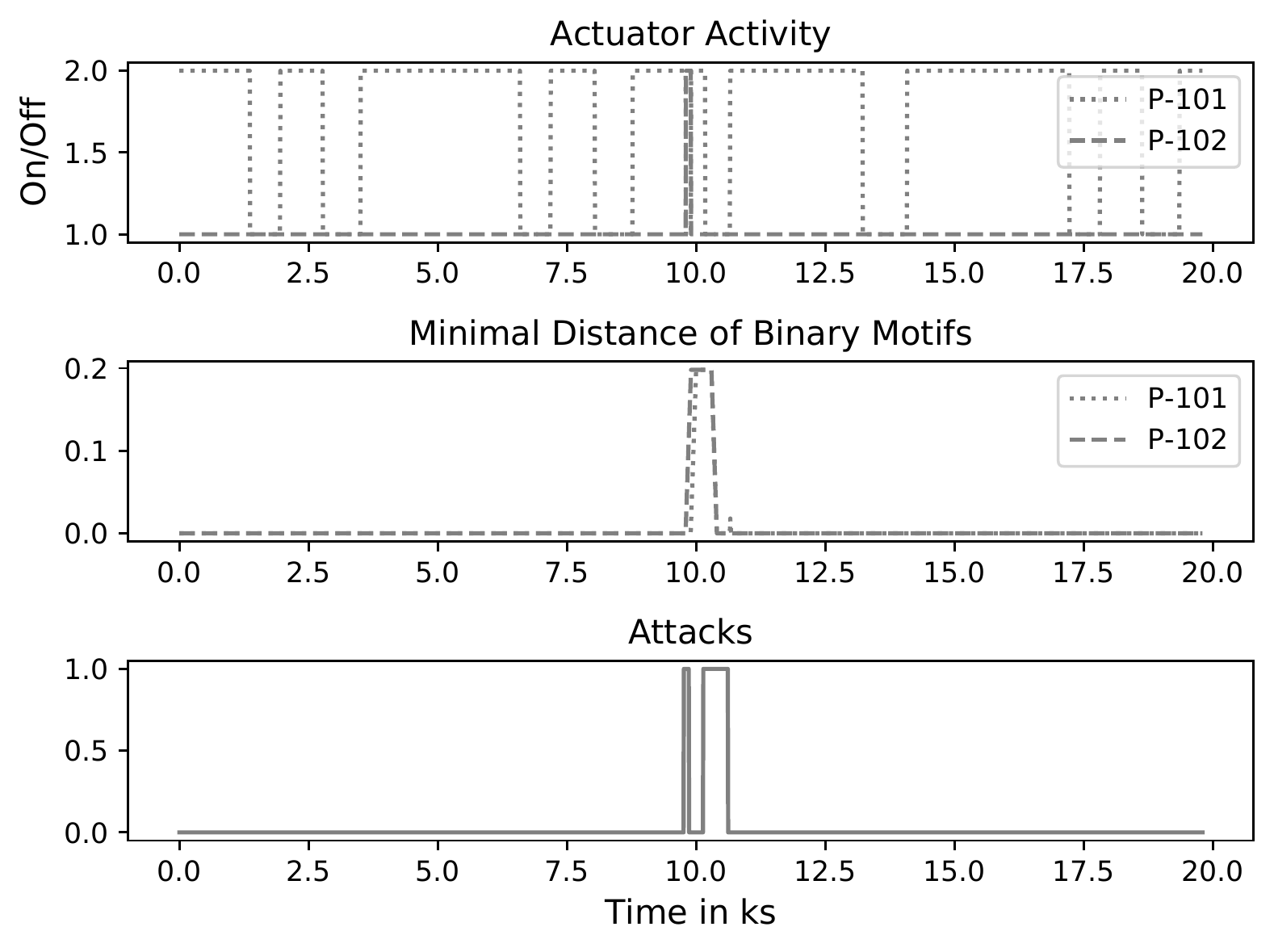}
\caption{\textit{Matrix Profile} of Interval 3 for \textit{P-101} and \textit{P-102}}
\label{fig:interval3}
\end{figure}
\begin{figure}[!ht]
\centering
\includegraphics[width=\linewidth]{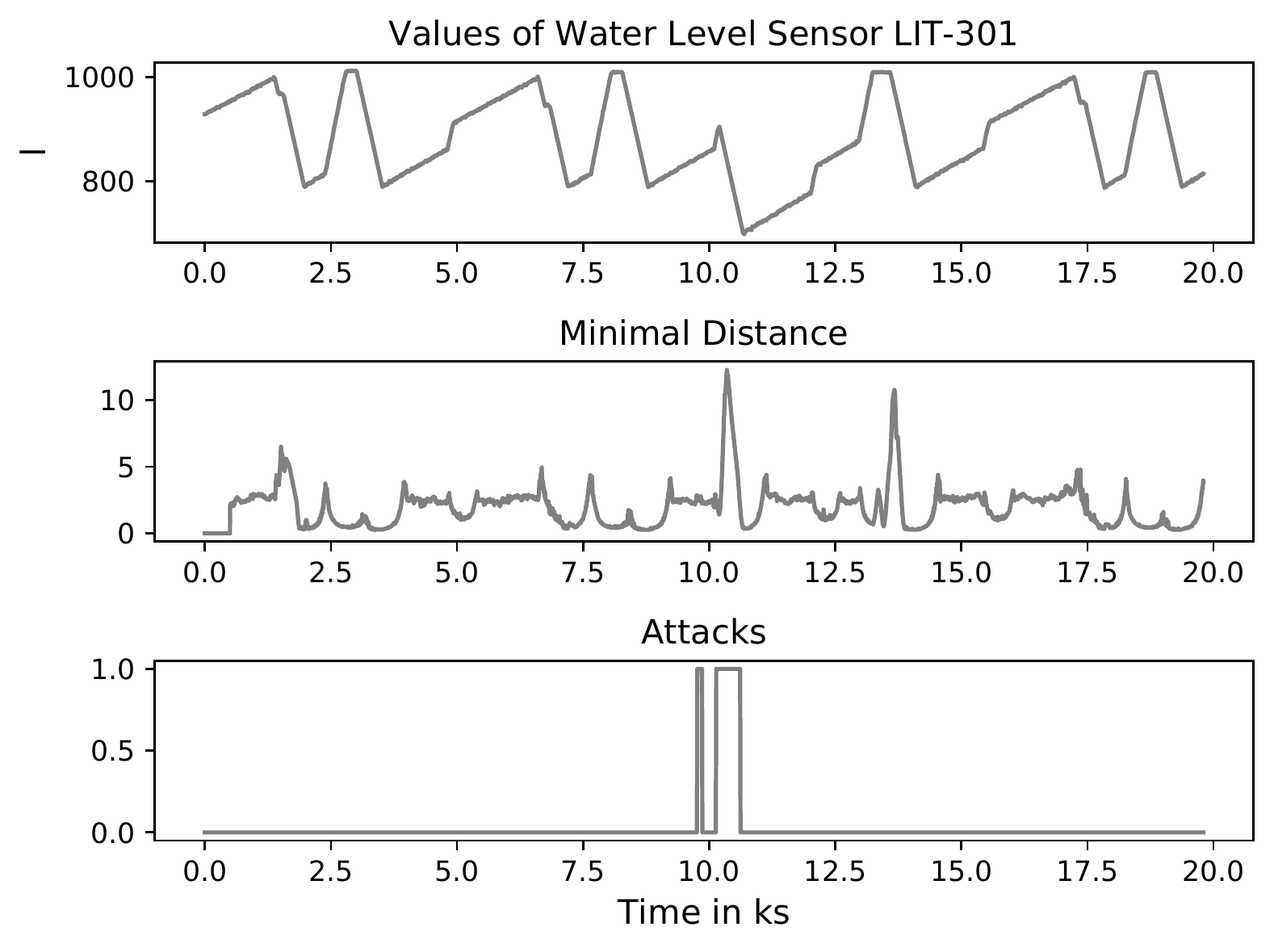}
\caption{\textit{Matrix Profile} of Interval 3 for \textit{LIT-301}}
\label{fig:int3_cont}
\end{figure}

Interval 3 is the period on January 1st  from 14:30:00 to 20:00:00, 
covering \numprint{19801} events during which two attacks occur.
The window size $m$ is set to \numprint{500},
as it produces better results than window sizes of \numprint{1000} and \numprint{2000}.
This contrasts the results of other works employing \textit{Matrix Profiles} for process data evaluation stating that the window size $m$ should not be smaller than the first peak of the autocorrelation function,
but might as well be larger~\cite{Duque_Anton.2018c}.
In interval 3,
the pump \textit{P-101} transporting water from the raw water tank into the second stage is observed.
For both attacks,
\textit{P-101} is turned off to stop the water flow.
However,
during the first attack,
backup pump \textit{P-102} is turned on so that the effect on the water level is not visible for the operator.
This behaviour of pumps \textit{P-101} and \textit{P-102} is shown in Figure~\ref{fig:interval3}.
The minimal distances of both actuators peak notably during the attacks.
Due to the length of the window size and the closeness of the attacks,
the minimal distance only contains one peak,
which starts at the beginning of the first attack.
The behaviour of the water level sensor \textit{LIT-301} during that period is shown in Figure~\ref{fig:int3_cont}.
The minimal distance for this sensor is calculated with the \ac{stamp} algorithm~\cite{Yeh.2016a}.
Due to the backup pump \textit{P-102},
the first attack cannot be detected by analysing the water level,
as there is no effect on the process,
except for an anomalous pump use.
In this case,
evaluating actuators can detect attacks that could not be detected otherwise.

\section{Discussion}
\label{sec:discussion}
The evaluation of the actuators provides an accurate detection of attacks on the process.
In case of interval 3,
the analysis of actuator behaviour is capable of detecting attacks that do not influence the system because of a backup pump \textit{P-102}.
Knowledge about an attempted attack that has failed is valuable for threat intelligence as it indicates the presence of an attacker.
Due to the uniform behaviour of actuators,
attacks are clearly distinguishable from normal operation by looking at the distance metrics.
It is typical for processes in batch processing to repeat for a large amount of time, 
providing a sound normal behaviour that is used as the basis for intrusion detection.
An advantage of \textit{Matrix Profiles} is the unsupervised fashion to deploy it.
In contrast to many machine learning-based approaches,
neither training nor labelled data are necessary to create sound models.
Furthermore,
it requires one hyper-parameter that is relatively robust.
Still,
choosing a window size $m$ is the most difficult task in employing \textit{Matrix Profiles}
Values that are too small lead to an increase in false negatives,
while window sizes that are too large lead to false positives.
Similar to previous works,
window sizes around the period length seem to be an optimal choice.

\section{Conclusion}
\label{sec:conclusion}
This work showed that the extended \textit{Matrix Profiles} employing the \textit{Hamming}-distance are capable of detecting all evaluated attacks without any false positives.
Furthermore,
training was performed automatically.
For future work,
extending the algorithm to extract motifs could prove beneficial in terms of computation time.
Since many sequences are expected to be identical,
a dictionary with a motif and its number of occurrences reduces the amount of comparisons required,
while providing the same amount of information.
Furthermore,
the number of occurrences can be used as an additional metric for an outlier,
as this would detect an attack that was deployed twice.
The second time would not be detected by regular \textit{Matrix Profiles} as the exact same motif was already present.
However,
any motif with a comparably low number of occurrences could be considered suspicious,
similar to the work of \textit{Duque Anton et al.}~\cite{Duque_Anton.2019g}.

\section*{Acknowledgement}
This work has been supported by the Federal Ministry of Education and Research of the Federal Republic of Germany (Foerderkennzeichen 16KIS0932, IUNO Insec).
The data set used in this work has been provided by \textit{iTrust, Centre for Research in Cyber Security, Singapore University of Technology and Design}.
The authors alone are responsible for the content of the paper.

\bibliographystyle{IEEEtran}
\bibliography{bibliography}

\end{document}